\documentclass[twocolumn,showpacs,preprintnumbers,amsmath,amssymb]{revtex4}

\usepackage{graphicx}
\usepackage{dcolumn}
\usepackage{bm}
\def\vector#1{{\boldsymbol{#1}}}

\def\vk{{\vector k}}

\def\vq{{\vector q}}
\def\vQ{{\vector Q}}

\def\vR{{\vector R}}

\def\dps{\displaystyle}

\def\vsigma{\boldsymbol{\sigma}}

\def\Tc{T_{c}}

\def\TAF{T_{\it AF}}

\def\HP{H_{\it P}}

\def\hMF{h_{\it MF}}

\def\muB{{\mu_B}}

\def\Wc{{W_c}}

\def\kB{{k_B}}

\def\JK{J_{\it K}}

\def\hsp#1{\hspace{#1ex}}

\def\eq.#1{eq.~(\ref{#1})}
\def\refeq#1{(\ref{#1})}

\def\kFpara{k_{F\parallel}}

\newcommand\Equation[2]{
\begin{equation}\label{#1}
#2
\end{equation}
}

\begin{document}

\title{
Interlayer spin-singlet pairing induced by magnetic interactions \\
in an antiferromagnetic superconductor 
}


\author{Hiroshi Shimahara}



\affiliation{
Department of Quantum Matter Science, ADSM, Hiroshima University, 
Higashi-Hiroshima 739-8530, Japan
}


\date{Received 11 July 2004} 

\begin{abstract}
It is shown that interlayer spin-singlet Cooper pairing is induced 
by magnetic interactions in a metallic antiferromagnet 
of stacked conductive layers 
in which each layer is ferromagnetically polarized and 
they order antiferromagnetically in stacking direction. 
As a result, the antiferromagnetic long-range order and 
superconductivity coexist at low temperatures. 
It is shown that $\TAF > \Tc$ 
except for in a very limited parameter region 
unless $\TAF = 0$, 
where $\TAF$ and $\Tc$ denote the antiferromagnetic and superconducting 
transition temperatures, respectively. 
It is found that the exchange field caused by 
the spontaneous staggered magnetization does not 
affect superconductivity at all, even if it is very large. 
The resultant superconducting order parameter has 
a horizontal line node, and is isotropic in spin space 
in spite of the anisotropy of the background magnetic order. 
We discuss the possible relevance of the present mechanism 
to the antiferromagnetic heavy fermion superconductors 
${\rm UPd_2Al_3}$ and ${\rm CePt_3Si}$. 
\end{abstract}


\maketitle


In this paper, we show that interlayer spin-singlet Cooper pairing 
is induced by magnetic interactions in 
a certain kind of 
metallic antiferromagnet. 
We consider a layered system of itinerant electrons in which 
each layer is ferromagnetically polarized 
but the majority-spin alternates in stacking direction. 
Therefore, the magetic order is characterized by the wave vector 
${\vector Q} = (0,0,\pi/c)$, 
where we have assumed the $a$ and $b$ crystal axes 
to be parallel to the layers, 
and the $c$-axis in the stacking direction, 
and $c$ denotes the $c$-axis lattice constant. 
It is also shown that 
the exchange field caused by spontaneous staggered magnetization 
does not influence superconductivity, however large it is.

The heavy fermion superconductors, 
such as ${\rm UPd_2Al_3}$ and ${\rm CePt_3Si}$, 
can be candidates of the present mechanism. 
The antiferromagnetic long-range order is considered 
to be characterized by the wave vector ${\vector Q} = (0,0,\pi/c)$, 
both in ${\rm UPd_2Al_3}$~\cite{Kit94} 
and in ${\rm CePt_3Si}$~\cite{Met04}. 
Superconducting transitions have been observed 
at $\Tc = 2.0$\,K and 0.7\,K, 
below the antiferromagnetic transition temperatures $\TAF = 14.3$\,K and 2.2\,K, 
in ${\rm UPd_2Al_3}$~\cite{Gei91}
and ${\rm CePt_3Si}$~\cite{Bau04}, respectively. 
It has been suggested that 
the magnetic moment is large, {\it i.e.}, $0.85\muB/{\rm U}$, 
in ${\rm UPd_2Al_3}$~\cite{Gei91}, 
but small, {\it i.e.}, $0.16\muB/{\rm Ce}$, 
in ${\rm CePt_3Si}$~\cite{Met04}.

The order parameter of interlayer spin-singlet pairing has 
a horizontal line node. 
This also agrees with the experimental results 
in the compound ${\rm UPd_2Al_3}$. 
The existence of the line node is suggested by 
the nuclear magnetic resonance (NMR) measurement~\cite{Tou95}, 
in which the Hebel-Slichter peak was absent, and $T_1^{-1} \propto T^3$ 
was observed. 
The singlet state is supported by the presence of 
NMR Knight shift~\cite{Tou95}, 
and the Pauli limited upper critical field~\cite{Ama92,Hes97}. 
In recent angle resolved magnetothermal transport 
measurements~\cite{Wat04}, 
two-fold oscillation in the rotation perpendicular 
to the $ab$ plane was observed, 
while no oscillation was observed 
in rotation in the $ab$ plane. 
These experimental results are reproduced 
by the order parameter of the form of 
$\Delta(\vk) = \Delta_0 \cos (k_z c)$.

Coexistence of superconductivity and magnetism has been studied 
in various models by many authors~\cite{And91,Kop66,Buz03}. 
In particular, 
spin singlet superconductivity 
in the presence of ferromagnetic layers 
has been studied by many authors~\cite{And91}. 
In the models examined in those papers, 
superconductivity occurs in a subsystem different from 
the magnetic layers. 
In contrast, we examine a model in which 
superconductivity occurs in electrons 
on magnetic layers in the present model. 
Kopaev also studied superconductivity 
when only magnetic electrons are present~\cite{Kop66}, 
although it exists only in the vicinity of the domain wall.

The magnetic structure mentioned above 
can be modeled most simply by the Hamiltonian 
\Equation{eq:H}
{
     H = H_0 + H_U + H_J 
     }
with the kinetic energy term 
\Equation{eq:H0}
{
     H_0 = \sum_{\vk\sigma} \xi({\vk}) \, 
               c_{\vk\sigma}^{\dagger} c_{\vk\sigma} , 
     }
the on-site Coulomb interactions 
\Equation{eq:HU}
{
     H_U = U \sum_{i} n_{i\uparrow} n_{i\downarrow} , 
     }
and the exchange interactions 
\Equation{eq:HJ}
{
     H_J = \frac{1}{2} \sum_{i,j} J_{ij}
               \bigl ( 
                 {\vector S}_i \cdot {\vector S}_j
                 - \frac{1}{4} n_i n_j
               \bigr ) . 
     }
We have defined 
${\vector S}_i = 
             \frac{1}{2} \sum_{\sigma \sigma'}
             c_{i\sigma}^{\dagger} \vsigma_{\sigma \sigma'} c_{i\sigma'}$, 
$n_{i} = \sum_{\sigma} n_{i\sigma}$, 
and $n_{i\sigma} = c_{i\sigma}^{\dagger} c_{i\sigma}$, 
where $\vsigma$ denotes the vector of Pauli matrices, 
and $c_{\vk\sigma}$ and $c_{i\sigma}$ denote 
the electron operators. 
We define $J_{ij} = J > 0$ for $\vR_j = \vR_i \pm {\hat c}$, 
$J_{ij} = -J_{\parallel} < 0$ for nearest neighbor sites $(i,j)$ 
on the same layer, 
and $J_{ij} = 0$ otherwise. 
Here, $\vR_i$ and ${\hat c}$ denote the lattice vector of the site $i$ 
and the unit lattice vector in the $c$-direction. 
The interlayer antiferromagnetic exchange interaction $J$ originates 
from the interlayer superexchange or kinetic exchange processes 
of electrons. 
The intralayer ferromagnetic exchange interaction $J_{\parallel}$ 
expresses the effect of the exchange Coulomb interaction, 
which is usually smaller than $U$ and $J$, 
but necessary to stabilize the present magnetic structure. 
The ferromagnetic correlation in each layer is due to 
$U$ and $J_{\parallel}$. 
However, two-dimensional (2D) long-range order without order 
in the $c$-direction cannot occur due to the thermal fluctuations 
in the present isotropic model, however large $U$ is. 
Transition to the long-range order occurs only in the presence of 
interlayer exchange interaction $J$, 
and the transition to the three dimensional antiferromagnetic 
long-range order at $T = \TAF$ is the only magnetic transition.

Many examples of compounds 
which can be modeled by 2D Heisenberg ferromagnetic model 
with antiferromagnetic interlayer exchange interactions are 
summarized by a review article by Jough and Miedema~\cite{Jou74}. 
For example, it was obtained from experimental data that 
$J/J_{\parallel} \approx 8 \times 10^{-3}$, 
$3.4 \times 10^{-3}$ and $0.21$, 
$\TAF = 13.8$, 16.8 and 18\,K, 
and 
$J_{\parallel}/\kB = 18.8$, 5.25 and 3.0\,K 
in the compounds ${\rm Rb_2CuCl_4}$, 
${\rm CrCl_3}$ and 
${\rm NaCrS_2}$, 
respectively. 
These compounds have the ferromagnetic short-range order in each layer 
at temperatures higher than the transition temperature 
due to the intralayer ferromagnetic exchange interaction $J_{\parallel}$, 
and undergo the long-range order 
by the weak interlayer exchange interaction $J$ 
at the transition temperature $\TAF$. 
In many examples, 
the interlayer exchange interactions are antiferromagnetic 
and much weaker than the intralayer interaction in most cases. 
In our model, we also take into account the on-site Coulomb repulsion $U$ 
in addition to the intralayer exchange interaction $J_{\parallel}$ 
to stabilize the ferromagnetic structure in each layer. 
Later, we consider a situation in which $U \gg J_{\parallel}$ as an example, 
but it is straightforward to apply the theory 
to the opposite case $U \ll J_{\parallel}$.

It is well-known due to the Mermin-Wagner's theorem~\cite{Mer66} 
that purely 2D isotropic Heisenberg model cannot exhibit 
any long-range order at any finite temperature. 
For the long-range order to be stabilized, 
an additional Ising type intralayer interaction 
or a three dimensional (interlayer) interaction 
is necessary. 
However, the former does not stabilize the present antiferromagnetic 
configuration in the stacking direction, as observed 
in ${\rm UPd_2Al_3}$~\cite{Kit94} and ${\rm CePt_3Si}$~\cite{Met04}. 
Therefore, more or less, interlayer antiferromagnetic interaction 
must exist in the present compounds. 
The physical origin of the antiferromagnetic interlayer interaction 
is interlayer kinetic exchange or superexchange process. 
In the former process, 
the interlayer exchange interaction $J$ is written 
as $J \sim 4 t_{\perp}^2/U$, where $t_{\perp}$ denotes interlayer 
electron hopping energy. 
Since $t_{\perp}$ is expected to be small from the crystal structure, 
the perturbation theory to derive the above experssion of $J$ 
would be justified.

When we apply the Hamiltonian \eq.{eq:H} to the compound ${\rm UPd_2Al_3}$, 
we should note that it has been suggested~\cite{Cas93,Gen92} 
by thermodynamic measurements 
that the magnetic and superconducting transitions 
occur in nearly disjunct subsystems in this compound. 
However, even if this is true, 
the present theory holds if the superconducting subsystem has 
a similar interlayer exchange interaction, 
which is plausible because both of the two subsystems have a $5f$ character 
and coexist in the same crystal structure. 
We discuss an application of the present theory taking into account 
the two-fluid model later.

The interaction terms can be rewritten as 
\Equation{eq:HUHJink}
{
     H_U + H_J = 
                 \frac{1}{N} \sum_{\vk \vk' \vq}
                 V(\vk,\vk',\vq) \, 
                 c_{\vk + \vq \uparrow}^{\dagger}     \, 
                 c_{\vk       \uparrow}               \, 
                 c_{\vk' - \vq \downarrow}^{\dagger}  \, 
                 c_{\vk'       \downarrow} 
     }
with 
\Equation{eq:Vkkq}
{
     \begin{split}
     V(\vk,\vk',\vq) 
              & = 
                   U - {\hat J}(\vq) - {\hat J}(\vk - \vk' + \vq) \\
              & \hsp{4}
                   + {\hat J}_{\parallel}(\vq) 
                   + {\hat J}_{\parallel}(\vk - \vk' + \vq) , 
     \end{split}
     }
where ${\hat J}(\vq) \equiv J \cos (\vq \cdot {\hat c}) $ 
and ${\hat J}_{\parallel}(\vq)$ are 
the Fourier transforms of the interlayer and intralayer 
exchange interactions, respectively. 
Since we have not specified the lattice structure of the layer, 
the expression of ${\hat J}_{\parallel}(\vq)$ is not shown, 
but it does not depend on $q_z$ 
and must have a peak around $\vq_{\parallel} = {\vector 0}$, 
where $\vq_{\parallel} = (q_x,q_y)$. 
Similarly, the form of $\xi(\vk)$ also depends 
on the lattice structure of the layer. 
We have simplified it as 
$
     \xi({\vk}) = \xi_{\parallel}(\vk_{\parallel}) + \xi_{\perp}(k_z) 
     $ 
with 
$
     \xi_{\parallel}(\vk_{\parallel}) 
       = {\hbar^2 |\vk_{\parallel}|^2}/{2 m^*} - \mu 
     $ 
and $\xi_{\perp}(k_z) \equiv - 2 t_{\perp} \cos (k_z c)$ 
for convenience, where $\vk_{\parallel} = (k_x,k_y)$. 
This simplification does not essentially change the qualitative results. 
In this paper, 
we examine the system in which $t_{\perp}$ and $J_{\parallel}$ are small. 
We take units with $\hbar = \kB = 1$.

First, we describe the magnetic transition. 
Let us examine the spin propagator 
\Equation{eq:chidef}
{
     \chi(\vR_i - \vR_j,\tau) 
          = - \langle T_{\tau} 
               S_{i}^{z}(\tau) S_{j}^{z} (0) \rangle 
     }
in the random phase approximation (RPA). 
We define the Fourier transform $\chi(\vq, i \nu_m)$ by 
\Equation{eq:chiFT}
{
     \chi(\vq, i \nu_m) 
       \equiv 
         \sum_{i} 
         \int_0^{\beta} \hsp{-1} d \tau \, 
            e^{-  i (\vq \cdot \vR_i - \nu_m \tau) } \, 
                 \chi(\vR_{i},\tau) , 
     }
where $\nu_m \equiv 2 \pi m T $ denotes the Matsubara frequency, 
and $\beta = 1/T$. 
If $\chi(\vQ,0)$ diverges, 
it indicates the phase transition to 
the magnetic long-range order with $\vQ$. 
If we omit $t_{\perp}$ and $J_{\parallel}$, we obtain 
\Equation{eq:chi}
{
     \chi(\vq, i \nu_m) 
     = \frac{1}{2} 
       \frac{\chi_0(\vq, i \nu_m)}
            { 1 - [ U - {\hat J}(\vq)] \, 
                  \chi_0(\vq, i \nu_m) } , 
     }
where 
\Equation{eq:chi0}
{
     \begin{split}
     \chi_0(\vq, i \nu_m) 
     & = - \frac{1}{N} 
     \sum_{\vk} T \sum_{n} 
         G_{\sigma}^{(0)} ( \vk,  i \omega_n ) 
     \\
     &    
     \hsp{4} 
     \times 
         G_{\sigma}^{(0)} ( \vk + \vq,  i \omega_n +  i \nu_m ) 
     \end{split}
     }
with the bare electron Green's function $G_{\sigma}^{(0)}$. 
When $t_{\perp} = 0$, the free susceptibility $\chi_0$ is 
expressed as 
\Equation{eq:chi0_2D}
{
     \chi_0(\vq,0) = \rho_0 \, 
     \left ( 
          1 - {\rm Re} \left [ 
               \sqrt{ 1 - \Bigl ( 
                              \frac{2 \kFpara}{q_{\parallel}}
                          \Bigr )^2} \, \, 
                       \right ] 
     \right ) 
     }
at $T = 0$, 
where $\rho_0 = m^{*} ab/2 \pi \hbar^2$ 
denotes the density of states of the 2D system. 
We have defined the effective mass $m^{*}$, 
the in-plane Fermi momentum $\kFpara$, 
and the lattice constants $a$ and $b$. 
The maximum $\chi(\vq, i \nu_m)$ occurs 
at arbitrary $\vq = (\vq_{\parallel},\pi/c)$ 
with $|\vq_{\parallel}| < 2 \kFpara$ 
and $\nu_m = 0$. 
This degeneracy is removed by $J_{\parallel} \ne 0$, 
which is small but exists in practice. 
Hence, $\chi(\vq, i \nu_m)$ reaches its maximum 
at $\vq = (0,0,\pi/c) \equiv \vQ$. 
It is easily verified by replacing $U - {\hat J}$ 
with $U - {\hat J} + {\hat J}_{\parallel}$ in \eq.{eq:chi}. 
Furthermore, when we take into account $t_{\perp} \ne 0$, 
$\chi_0$ has a peak around $\vq = (\vq_{\parallel},\pi/c) = \vQ'$ 
with $|\vq_{\parallel}| \approx 2 \kFpara$, 
but the difference $\chi_0(\vQ',0) - \chi_0({\vector 0},0) 
\propto t_{\perp}^2$ is small. 
Therefore, when $t_{\perp} \ne 0$, we must assume $J_{\parallel} \ne 0$ 
which is small but sufficiently large for the maximum of $\chi$ 
to occur at $\vq = \vQ$, so that the magnetic order of $\vQ$ 
is stabilized.

When these conditions are satisfied, 
antiferromagnetic transition occurs at a temperature 
which satisfies 
\Equation{eq:TAFcondition}
{
     1 = (U + J) \, \chi_0(\vQ,0) 
     }
from \eq.{eq:chi}, where 
$
     \chi_0(\vQ,0)
       = \chi_0({\vector 0},0) 
       = \rho_0 / (e^{- \beta \mu} + 1) 
     $. 
The chemical potential $\mu$ is determined by the equation for 
the electron number per site 
$
     n = 2 \beta^{-1} \rho_0 \ln (1 + e^{\beta \mu}) 
     $. 
Thus, we obtain 
$
     \chi_0(\vQ,0) = \rho_0 ( 1 - e^{- \beta n / 2 \rho_0} ) 
     $. 
Therefore, we obtain the antiferromagnetic transition temperature 
\Equation{eq:TAF}
{
     \TAF 
     = \frac{n}{\dps{ 2 \rho_0 
                      \ln \frac{(U + J) \rho_0}{(U + J) \rho_0 - 1}}} 
     }
to an ordered state with the wave vector 
$\vQ = (0,0,\pi/c)$ when $(U + J) \rho_0 > 1$, 
while $\TAF = 0$ otherwise.

In the antiferromagnetic phase, the electron states are affected by 
spontaneous staggered magnetization. 
We define $A$ and $B$ sublattices (sublayers) whose majority spins 
are up and down, respectively. 
We write the electron operators as $a_{i\sigma}$ and $b_{j\sigma}$ 
for $i \in A$ and $j \in B$. 
Therefore, we have 
\Equation{eq:na_and_nb}
{     
     \begin{split} 
     n^{A}_{i\sigma} 
       & = \langle a_{i\sigma}^{\dagger} a_{i\sigma} \rangle 
         = \frac{n}{2} + \sigma m , 
     \\
     n^{B}_{j\sigma} 
       & = \langle b_{j\sigma}^{\dagger} b_{j\sigma} \rangle 
         = \frac{n}{2} - \sigma m , 
     \end{split}
     }
where 
${
     \langle S_{i}^{z} \rangle = - \langle S_{j}^{z} \rangle = m  
     }$ 
for $i \in A$ and $j \in B$. 
We have defined $\sigma = + 1$ and $-1$ in equations, 
which correspond to $\sigma = \uparrow$ and $\downarrow$ in suffixes, 
respectively. 
Corrections to the kinetic energy due to $H_U + H_J$ 
are taken into account by the mean field approximation as 
\Equation{eq:HUHJ_MF}
{
     H_{\rm MF} 
          = - {\sum_{\vk\sigma}}' 
                \sigma \hMF 
                \bigl [ 
                  a_{\vk \sigma}^{\dagger} a_{\vk \sigma} 
                - b_{\vk \sigma}^{\dagger} b_{\vk \sigma} 
                \bigr ] 
     }
with $\hMF = (U + J + z_{\parallel} J_{\parallel}/2) \, m$, 
where $z_{\parallel}$ denotes the number of nearest neighbor sites 
in the layer, 
and the summation ${\sum}'_{\vk}$ is carried out 
over the half Brillouin zone. 
Therefore, the total kinetic energy term 
${\tilde H}_0 \equiv H_0 + H_{\rm MF}$ is 
written as 
\Equation{eq:H0HMF}
{
     \begin{split}
     {\tilde H}_0 
       & = {\sum_{\vk\sigma}}' 
            \Bigl [ 
                \xi_{\parallel \sigma}^{A} 
                   a_{\vk \sigma}^{\dagger} a_{\vk \sigma} 
              + \xi_{\parallel \sigma}^{B} 
                   b_{\vk \sigma}^{\dagger} b_{\vk \sigma} 
       \\[-4pt] 
       & \hsp{8} 
              + \xi_{\perp}
                 \bigl ( 
                   a_{\vk \sigma}^{\dagger} b_{\vk \sigma} 
                 + b_{\vk \sigma}^{\dagger} a_{\vk \sigma} 
                 \bigr ) 
            \Bigr ] , 
     \end{split}
     }
where 
$\xi_{\parallel \sigma}^{A} = \xi_{\parallel} - \sigma \hMF$ 
and 
$\xi_{\parallel \sigma}^{B} = \xi_{\parallel} + \sigma \hMF$. 
The mean field approximation is consistent with the RPA, 
which we have used to derive $\TAF$, 
although we have neglected $J_{\parallel}$ in \eq.{eq:TAF}.

Now, let us examine superconductivity. 
We will show its formulation in the antiferromagnetic phase, 
but it is immediately reduced to that in the paramagnetic phase 
by putting $m = 0$. 
The exchange interaction $H_J$ contributes to pairing interaction, 
while it causes the antiferromagnetic transition 
and creates the exchange field. 
We rewrite $H_J$ as 
\Equation{eq:HJPsiPsisinglet}
{
     H_J = - \hsp{-2}
             \sum_{i \in A, j \in B} 
             \hsp{-2}
             J_{ij}   {\Psi_{ij}^{\rm (s)}}^{\dagger}
                       \Psi_{ij}^{\rm (s)} , 
     }
where 
${
     \Psi_{ij}^{\rm (s)}
     = 2^{-1/2}
     (   a_{i\uparrow} b_{j\downarrow} 
                            - a_{i\downarrow} b_{j\uparrow} ) 
     }$. 
The statistical average $\langle \Psi_{ij}^{\rm (s)} \rangle$ 
is the order parameter of interlayer spin-singlet pairing. 
Here, we have neglected $J_{\parallel}$, since it does not have 
an important effect on superconductivity if it is small. 
In \eq.{eq:HJPsiPsisinglet}, 
it is found that $J$ contributes only to spin-singlet pairing 
as an attractive interaction in its first order. 
In the BCS approximation, \eq.{eq:HJPsiPsisinglet} is written as 
\Equation{eq:HJinDelta}
{
     H_J \approx \sum_{\vk \sigma} \sigma 
             \bigl [ 
               \Delta (\vk) \, a_{\vk \sigma} b_{- \vk - \sigma}
             + {\it h.c.} 
             \bigr ] 
     }
with the order parameter 
\Equation{eq:Deltadef}
{
     \Delta(\vk) 
     = - \frac{1}{2N} 
         {\sum_{\vk'\sigma'}}' 
         \sigma' {\hat J}(\vk - \vk') \, 
           \langle 
                   b_{- \vk'  - \sigma'}^{\dagger} 
                   a_{\vk' \sigma'}^{\dagger} 
           \rangle . 
     }
Therefore, we obtain 
\Equation{eq:Deltaresult}
{
     \Delta(\vk) = \Delta_0 \cos (k_z c)
     }
with 
\Equation{eq:Delta0def}
{
     \Delta_0 
     = - \frac{J}{2N} 
         {\sum_{\vk'\sigma'}}' 
         \sigma' \cos (k'_z c)
           \langle 
                   b_{- \vk'  - \sigma'}^{\dagger} 
                   a_{\vk' \sigma'}^{\dagger} 
           \rangle . 
     }
In the same approximation, the on-site Coulomb interaction $H_U$ 
is ineffective for anisotropic superconductivity.

It has been proposed that in the higher order of $J$, 
the pairing interaction is enhanced 
by the exchange of magnons~\cite{Shi94}. 
This mechanism has also been examined 
in the compound ${\rm UPd_2Al_3}$~\cite{Sat01,NoteSat01,Miy01,NoteMiy01}. 
However, since the spin fluctuations are weak 
at temperatures much lower than 
the antiferromagnetic transition temperature 
($T \lesssim \TAF/7$), 
the pairing interaction mediated by the magnons is weak~\cite{Shi94}. 
Hence, in this paper, we neglect them in comparison to the direct 
pairing interaction described 
in eqs.~\refeq{eq:HJPsiPsisinglet} and~\refeq{eq:HJinDelta}. 
The present direct pairing interaction is not mediated by the magnons. 
In a broader sense, however, 
one may regard the present pairing interaction as 
mediated by the spin fluctuations, 
because the superexchange interactions are derived from 
the virtual process of electrons, with which their spins correlate.

When $J = 0$, the present model is reduced to 
the quasi-2D Hubbard model. 
In the perturbation theory based on it, 
it was shown that antiferromagnetic fluctuations induce 
intralayer singlet pairing 
near the antiferromagnetic phase~\cite{Shi88}. 
However, the present system is ferromagnetic in each layer. 
In the absence of $J$, 
the propagator of the fluctuations $\chi(\vq)$ 
has a broad peak around 
$\vq_{\parallel} \approx {\vector 0}$ 
as we can see in \eq.{eq:chi}, 
in contrast to the sharp peak in the antiferromagnetic case, 
where the Fermi-surface nesting occurs. 
Therefore, the pairing interaction is not strongly enhanced by 
the spin fluctuation unless $U$ is large. 
Nisikawa and Yamada~\cite{Nis02} 
examined the ${\rm UPd_2Al_3}$ on the basis of 
a 2D Hubbard model taking into account the lattice structure, 
although they did not examine interlayer pairing. 
In the presence of interlayer interactions, $t_{\perp}$ and $J$, 
the spin fluctuations will enhance 
the interlayer singlet pairing interaction.

From eqs.~\refeq{eq:H0HMF} and~\refeq{eq:HJinDelta}, 
we obtain 
\Equation{eq:HMF}
{
     H = \sum_{\vk} \psi_{\vk}^{\dagger} {\hat M} \psi_{\vk} , 
     }
where we have defined 
${
     \psi_{\vk}^{\dagger} \equiv 
         \bigl ( 
                 a_{\vk \uparrow}^{\dagger} \hsp{1} 
                 b_{\vk \uparrow}^{\dagger} \hsp{1} 
                 a_{- \vk \downarrow} \hsp{1} 
                 b_{- \vk \downarrow} 
         \bigr ) 
     }$ 
and 
\Equation{eq:Mhatdef}
{
     {\hat M}(\vk)
       \equiv 
         \left ( \begin{array}{cccc} 
                 \xi_{\parallel \uparrow}^{A} 
                   & \xi_{\perp} & 0 & - \Delta^{*} 
                 \\[4pt]
                 \xi_{\perp} & \xi_{\parallel \uparrow}^{B} 
                   & - \Delta^{*} & 0 
                 \\[4pt]
                 0 & - \Delta & - \xi_{\parallel \downarrow}^{A} 
                   & - \xi_{\perp} 
                 \\[4pt]
                 - \Delta & 0 & - \xi_{\perp} 
                   & - \xi_{\parallel \downarrow}^{B} 
                 \end{array}
         \right ) . 
         }
The Green's function is defined in matrix form by 
\Equation{eq:Ghatdef}
{
     {\hat G}(\vk,\tau) 
        = - \langle T_{\tau} 
            \psi_{\vk}(\tau) \psi_{\vk}^{\dagger} \rangle . 
     }
From the equation of motion, we obtain 
\Equation{eq:Ghatresult}
{
     {\hat G}(\vk, i \omega_n) 
          = \bigl [
                i \omega_n {\hat I} - {\hat M}(\vk) 
            \bigr ]^{-1} ,
     }
where ${\hat I}$ denotes the $2 \times 2$ unit matrix. 
We obtain the quasi-particle energies, 
$\pm E_A(\vk)$ and $\pm E_B(\vk)$, 
where 
$E_{A}^2 = E^2 + F$, $E_{B}^2 = E^2 - F$, 
$ E^2 = \xi_{\parallel}^2 + \xi_{\perp}^2 + \hMF^2 + |\Delta|^2$ 
and 
$ F = 2 [ \xi_{\parallel}^2 
                      ( \xi_{\perp}^2 + \hMF^2 ) 
                    + {|\Delta|^2 \xi_{\perp}^2}/{2}]^{1/2}$. 
If we define $G_{ij}$ as the $(i,j)$ component of the matrix ${\hat G}$, 
we have 
\Equation{eq:langle_ba_rangle}
{
     \begin{split}
     \langle 
            b_{\vk'   \uparrow}^{\dagger} 
            a_{- \vk' \downarrow}^{\dagger} 
     \rangle 
     & = G_{32}(\vk', \tau = - 0) , 
     \\
     \langle 
            b_{- \vk' \downarrow}^{\dagger} 
            a_{\vk'   \uparrow}^{\dagger} 
     \rangle 
     & = - G_{41}(\vk', \tau = - 0) . 
     \end{split}
     }
Therefore, we obtain the gap equation 
\Equation{eq:gapeq_final}
{
     \Delta(\vk) 
       = \frac{1}{4 N} {\sum_{\vk'}}' 
             {\hat J}(\vk - \vk') \hsp{-1} 
             \sum_{X = A,B} \hsp{-1}
               \frac{\tanh \frac{E_{X}(\vk')}{2T}}
                    { 2 E_{X}(\vk') } 
       \Delta(\vk') , 
     }
In the limit of $\Delta_0 \rightarrow 0$, 
we have $E_{X} = | \xi_{X} |$, 
where 
$\xi_{A} = \xi_{\parallel} + \delta_{\perp}$, 
$\xi_{B} = \xi_{\parallel} - \delta_{\perp}$, 
and $\delta_{\perp} = \sqrt{\hMF^2 + \xi_{\perp}^2}$. 
Therefore, we obtain the equation for $\Tc$ 
\Equation{eq:Tc}
{
     \begin{split}
     1 & = \frac{J}{4 N} {\sum_{\vk'}}' 
             \cos^2 (k_z' c) 
             \sum_{X = A,B} 
               \frac{\tanh [{\xi_{X}(\vk')}/{2 \Tc }] }
                    { 2 \xi_{X}(\vk') } . 
     \end{split}
     }
It is easily verified that the right hand side of 
\eq.{eq:Tc} exhibits a logarithmic divergence 
in the limit of $T \rightarrow 0$, 
irrespective of the value of the exchange field $\hMF$. 
Logarithmic divergence, 
which results from the Fermi-surface instability, 
is crucial for the occurrence of superconductivity. 
Since the divergence occurs 
even if $\hMF$ is very large, 
the existence of a large local magnetic moment as 
observed in ${\rm UPd_2Al_3}$~\cite{Gei91} 
does not deny the occurrence of superconductivity 
in the electrons which have the present magnetic structure. 
It is interesting that superconductivity is not influenced 
by large spin polarization in each layer, 
although such polarization creates the strong exchange field 
on the electrons responsible to superconductivity. 
It is well-known 
that in ferromagnets, strong exchange field suppresses 
superconductivity by Pauli paramagnetic pair-breaking effect.

\begin{figure}
\vspace{2ex} 
\vspace{2ex} 
\begin{center}
\includegraphics[width=6.0cm]{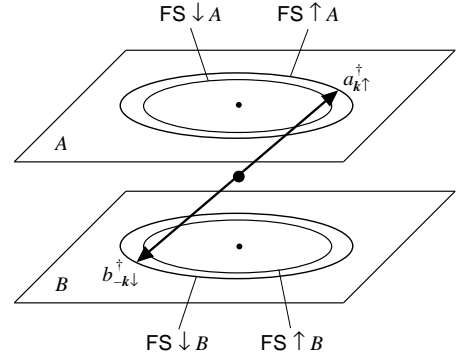}
\end{center}
\caption{
Schematic figure of the Fermi surfaces of 
$\uparrow$ and $\downarrow$ electrons 
in the $A$ and $B$ layers, 
and interlayer spin-singlet pairing. 
The abbreviation FS$\sigma \! X$ denotes 
the Fermi surface of the spin $\sigma$ electrons on the $X$ sublattice, 
where $\sigma = \uparrow$, $\downarrow$ 
and $X = A$, $B$. 
The splits of the Fermi surfaces of the $\uparrow$ and $\downarrow$ 
spin electrons do not affect interlayer spin-singlet pairing. 
} 
\label{fig:FSAB}
\end{figure}

This result can easily be verified as follows. 
The present superconductivity is due to singlet pairing of 
$a_{\vk \sigma}$ and $b_{-\vk -\sigma}$ electrons, 
{\it i.e.}, spin $\sigma$ electrons on $A$ sublattice 
and spin $- \sigma$ electrons on $B$ sublattice. 
When we define the Fermi surface of each sublattice, 
the magnitudes of the Fermi momenta of those electrons are equal, 
irrespective of the magnitude of the Zeeman splitting 
due to the exchange field in each layer, 
as schematically shown in Fig.~\ref{fig:FSAB}. 
Therefore, the present pair states are not influenced 
by the magnetic moments.

Needless to say, interlayer pairing does not mean that 
the coherence length in the $c$-direction $\xi_{0 \perp}$ is 
on the order of the layer spacing. 
In the present system, $\xi_{0 \perp}$ is 
on the order of $v_{F \perp}/\Delta_0$, 
where $v_{F \perp}$ and $\Delta_0$ denote Fermi velocity 
in the $c$-direction and the scale of magnitude of 
the order parameter at $T=0$, respectively. 
We obtain $\xi_{0\perp} \gg c$, if $t_{\perp} \gg \Delta_0$.

When $t_{\perp}$ is negligible, 
the transition temperature $\Tc$ is obtained as follows. 
Since $\delta_{\perp} = |\hMF|$, 
we can integrate $\cos^2(k_z'c)$ with respect to $k_z'$ first 
in \eq.{eq:Tc}. 
Unless $\delta_{\perp} > \mu$, the density of states is constant 
when $t_{\perp} \approx 0$. 
The $\vk_{\parallel}$ integral is 
approximated by the $\xi_{\parallel} \pm \hMF$ integrals 
with an effective cutoff energy $\Wc$, 
which is on the order of the band width. 
More explicitly, it is expressed as 
$\Wc = [(W-\mu-\hMF)(W-\mu+\hMF)(\mu+\hMF)(\mu-\hMF)]^{1/4}$, 
where $W$ and $\mu$ denote the band width and the chemical potential 
measured from the bottom of the band, respectively. 
Carrying out the integral, we obtain 
\Equation{eq:Tcresult}
{
     \Tc = 1.13 \, \Wc \, e^{ - 2/J\rho_0 } . 
     } 
Here, it is found that $\Tc$ is not influenced 
by the spontaneous staggered magnetization $g \muB m$, 
as expected from the above argument.

\begin{figure}
\vspace{2ex} 
\vspace{2ex} 
\begin{center}
\includegraphics[width=7.0cm]{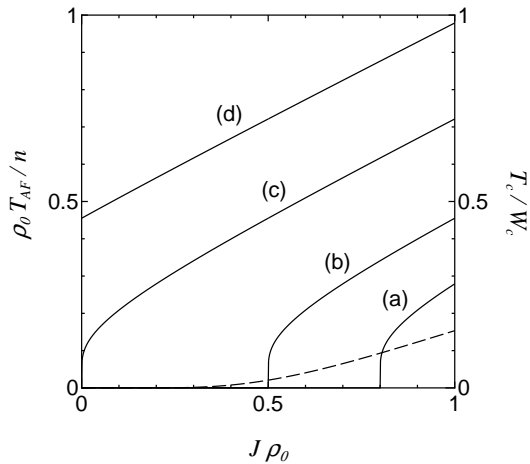}
\end{center}
\caption{
The transition temperatures as functions of $J \rho_0$. 
The solid curves (a) - (d) show the results of $\rho_0\TAF/n$ 
for $U\rho_0 = 0.2$, 0.5, 1.0, and 1.5, respectively. 
The dashed curve shows the results of $\Tc/\Wc$. 
} 
\label{fig:TAFTc}
\end{figure}

Figure~\ref{fig:TAFTc} depicts 
the antiferromagnetic transiton temperature $\TAF$ 
and the superconducting transition temperature $\Tc$ 
scaled by $n/\rho_0$ and $\Wc$, respectively. 
These scales have the same orders of magnitude, 
{\it i.e.}, $\rho_0 \sim 1/\Wc$. 
It is found that, as the order of the magnitudes, 
the experimental results in ${\rm UPd_2Al_3}$, 
$\TAF = 14.3$\,K and $\Tc = 2.0$\,K could be explained within 
the present mechanism, if the effective band width is on the order of 
$W = 10 \sim 100$\,K, which is realistic 
for heavy fermion systems.

\begin{figure}
\vspace{2ex} 
\vspace{2ex} 
\includegraphics[width=6.0cm]{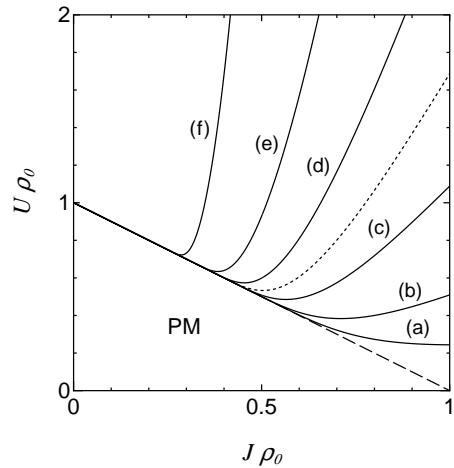}
\caption{
Contour lines for various ratios $\TAF/\Tc$ in $J$-$U$ plane. 
The solid curves (a) - (f) are the countour lines for 
$\alpha = 2$, 3, 5, 10, 20, and 100, respectively, 
where $\alpha \equiv 
(\TAF\rho_0/n)/(\Tc/\Wc) = \TAF/\Tc \times \Wc\rho_0/n$. 
The dotted curve is that of $\alpha = 7$, 
which corresponds to the experimental result of ${\rm UPd_2Al_3}$. 
The dashed curve shows the phase boundary 
between the antiferromagnetic and paramagnetic (PM) phases 
at $T = 0$. 
}
\label{fig:UJphased}
\end{figure}

Figure~\ref{fig:UJphased} shows the values of $J\rho_0$ and $U\rho_0$ 
for a given ratio of 
$\alpha \equiv (\TAF\rho_0/n)/(\Tc/\Wc) = \TAF/\Tc \times \Wc\rho_0/n$. 
If we consider $\Wc\rho_0/n \sim 1$, 
the experimental data $\TAF = 14.3$\,K and $\Tc = 2.0$\,K 
give $\alpha \sim 7$. 
Therefore, the experimental value of ratio $\alpha$ can be reasonably 
reproduced for moderate values of the coupling constants $J$ and $U$. 
We will compare the theoretical and experimental results 
more closely below. 
We obtain $\TAF > \Tc$ or $\TAF = 0$, 
except in a very small region of the phase diagram.

We note that the resultant singlet order parameter is 
invariant under rotation in spin space. 
The rotational transformation is made by 
\Equation{eq:Rtheta}
{
     R(\theta) = \exp \bigl [ i \frac{\theta}{2} \sigma_y \bigr ] 
       = \cos \frac{\theta}{2} + i \sigma_y \sin \frac{\theta}{2} . 
     }
The electron operators in the rotated space are defined by 
\Equation{eq:ctilde}
{
     \left ( 
     \begin{array}{c} 
      {\tilde c}_{\vk \uparrow}^{\dagger} 
      \\[4pt]
      {\tilde c}_{\vk \downarrow}^{\dagger}
     \end{array}
     \right ) 
     = 
     R(\theta) 
     \left ( 
     \begin{array}{c}
      c_{\vk \uparrow}^{\dagger}
      \\[4pt]
      c_{\vk \downarrow}^{\dagger}
     \end{array}
     \right ) , 
     }
where $c_{\vk \sigma} = a_{\vk \sigma}$ or $b_{\vk \sigma}$. 
It is easily verified that 
\Equation{eq:singletrotated}
{
     \sum_{\sigma} \sigma 
       \langle b_{-\vk -\sigma}^{\dagger}
               a_{\vk \sigma}^{\dagger} \rangle 
     = 
     \sum_{\sigma} \sigma 
       \langle {\tilde b}_{-\vk -\sigma}^{\dagger}
               {\tilde a}_{\vk \sigma}^{\dagger} \rangle . 
     } 
In addition, the present pairing interaction is rotationally invariant, 
{\it i.e.}, ${\hat J}(\vk,\vk')$ does not have spin suffixes. 
Therefore, $\Delta(\vk)$ given by \eq.{eq:Deltadef} 
is isotropic in spin space, 
irrespective of the direction of the magnetic order. 
In fact, the resultant gap equations~\refeq{eq:gapeq_final} 
and~\refeq{eq:Tc} are invariant under spin rotation.

These results may explain the experimental data of 
the muon spin rotation measurements in ${\rm UPd_2Al_3}$~\cite{Fey94}, 
in which it was observed that the London penetration depth 
and the magnetic susceptibility reduction below $\Tc$ 
are essentially isotropic, 
while the total susceptibility remains strongly anisotropic. 
In the present mechanism, 
not only the singlet nature but also the rotational invariance 
of the pairing interaction play essential roles in the spin isotropy 
of $\Delta(\vk)$. 
In contrast, in the ``magnetic exciton'' mechanism, 
the pairing interaction and the order parameter are anisotropic 
if the magnetic system is anisotropic.

The compound ${\rm CePt_3Si}$ is another candidate for 
the present pairing mechanism, 
although the ratio of the transition temperatures $\TAF/\Tc$ 
is much smaller than that in ${\rm UPd_2Al_3}$. 
In contrast to ${\rm UPd_2Al_3}$, the critical field largely exceeds 
the Pauli paramagnetic limit $\HP$ estimated by the simplified formula 
$H_{\rm P} \approx 1.86 [{\rm T/K}] \times T_{\rm c} [{\rm K}] 
\approx 1.4 {\rm T}$ in ${\rm CePt_3Si}$~\cite{Bau04}. 
If the present mechanism of singlet pairing 
is realized in ${\rm CePt_3Si}$, 
the large critical field cannot be attributed to equal spin pairing. 
It can be explained by an effect of exchange field created by 
coexisting antiferromagnetic long-range order, 
which reduces the Pauli paramagnetic pair-breaking effect~\cite{Shi04}. 
It is still controversial whether dominant pairing in ${\rm CePt_3Si}$ 
is of singlet or triplet. 
Recent NMR data suggests that the gap function may have 
some novel structure~\cite{Yog04}. 
Thermal transport measurements have suggested that 
the order parameter has line nodes~\cite{Iza05}, 
which is consistent with the present theory. 
It is known that the compound ${\rm CePt_3Si}$ does not have 
inversion symmetry. 
The Rashba interaction~\cite{Ras60} 
has been examined to include it. 
Yip predicted that 
the Knight shift vanishes 
in the superconductors with strong Rashba interaction.~\cite{Yip02}

For more close comparison with the experimental data of ${\rm UPd_2Al_3}$, 
we consider a two-fluid model~\cite{Cas93,Gen92}. 
In order to take into account the model, 
we assume two different renormalization factors $Z_s$ and $Z_m$ 
for electrons responsible to superconductivity 
and antiferromagnetic order, respectively. 
We should note that in actuality 
there are strong renormalization effects 
in the heavy fermion system, 
not only in the density of states, but also in the vertex corrections. 
It is easily verified by diagramatical consideration 
that in terms of $Z_s$ and $Z_m$, 
the electron mass $m_0$, the density of states $\rho_0$, 
the band width $W$, 
and the coupling constants $J$ and $U$ are modified as 
${\tilde m}_a = Z_a m_0$, 
${\tilde \rho}_a = Z_a \rho_0$, 
${\tilde W}_a = W/Z_a$ 
${\tilde J}_a = J/Z_a^2$ 
and ${\tilde U}_a = J/Z_a^2$, 
respectively, 
where $a = s, m$. 
The experimental value $\gamma_1 = 115 {\rm mJ/K^2mol}$~\cite{Cas93} 
gives ${\tilde \rho}_s = 1/476\,{\rm K}^{-1}$. 
Equation~\refeq{eq:Tcresult} is rewritten as 
$\Tc = 1.13 \Wc e^{-1/{\tilde \lambda}}$ 
with ${\tilde \lambda} = {\tilde J}_s {\tilde \rho}_s$, 
$\rho_s \sim 1/{\tilde W}_s$, 
and $\Wc = {\tilde W}_s/2$. 
Hence, $\Tc = 2.2$\,K gives ${\tilde J}_s = 173\,{\rm K}$. 
The antiferromagnetic transition temperature of \eq.{eq:TAF} 
is written as 
\Equation{eq:TAFrenormalized}
{
     \TAF 
     = \frac{n}{\dps{ 2 {\tilde \rho}_m 
                        \ln \frac{({\tilde U}_m + {\tilde J}_m) 
                                  {\tilde \rho}_m }
                                 {({\tilde U}_m + {\tilde J}_m) 
                                  {\tilde \rho}_m - 1}}} . 
     }
Here, we simply put $n = 2$ for order estimations. 
From the values of ${\tilde J}_s$ and ${\tilde \rho}_s$ estimated above, 
the values of ${\tilde J}_m$ and ${\tilde \rho}_m$ can be obtained, 
if the ratio $Z_m/Z_s$ is known. 
Therefore, we only need the values of $Z_m/Z_s$ and ${\tilde U}_m$ 
for estimation of $\TAF$. 
However, since they are not known for ${\rm UPd_2Al_3}$ at the present, 
we need to assume them. 
In the assumption, 
we require that ${\tilde W}_m$ is smaller than $\TAF$ 
consistently with the observation of a large local magnetic moment 
in ${\rm UPd_2Al_3}$~\cite{Cas93}. 
Physically, it is also plausible 
that ${\tilde U}_m$ is not much larger than $\TAF$ 
but larger than ${\tilde W}_m$. 
As an example, let us assume that $Z_m/Z_s = 30$, 
which gives ${\tilde W}_m \approx 8\,{\rm K} \ll \TAF$ 
and ${\tilde J}_m \approx 0.19\,{\rm K}$. 
In this case, 
if we assume ${\tilde U}_m = 20\,{\rm K}$ and $30\,{\rm K}$ as examples, 
we obtain $\TAF \approx 10\,{\rm K}$ and $21\,{\rm K}$, respectively. 
As another example, we assume $Z_m/Z_s = 20$, 
which gives ${\tilde W}_m \approx 12\,{\rm K} \ll \TAF$ 
and ${\tilde J}_m \approx 0.43\,{\rm K}$. 
In this case 
${\tilde U}_m = 30\,{\rm K}$ and $40\,{\rm K}$ give 
$\TAF \approx 16\,{\rm K}$ and $27\,{\rm K}$, 
respectively. 
The values ${\tilde J}_m \approx 0.19\,{\rm K}$ and 
$0.43\,{\rm K}$ obtained in these examples are not outrageous 
as energy parameters in real materials. 
In fact, $J \approx 0.15\,{\rm K}$ and $0.63\,{\rm K}$ were 
obtained from experimental data 
in ${\rm Rb_2CuCl_4}$ and ${\rm NaCrS_2}$, 
respectively~\cite{Jou74}. 
Here, $J$ was not estimated as the bare parameter 
but estimated as the dressed (observed) parameter 
like ${\tilde J}_m$. 
Although our estimations are crude, 
the values of $\TAF$ obtained above are on the same order 
of the experimental result $\TAF = 14.3$\,K. 
Therefore, we find that 
the present mechanism reproduces consistent orders of 
magnitudes of $\Tc$ and $\TAF$ 
for appropriate values of $Z_m/Z_s$ and ${\tilde U}_m$.

The result that $\TAF > \Tc$ except for in a very limited region 
unless $\TAF \ne 0$ can be explained physically as follows. 
In the present mechanism, both interlayer antiferromagnetic 
long-range order and interlayer singlet superconductivity 
are induced by interlayer antiferromagnetic exchange interaction. 
However, strong on-site repulsion contributes to stabilization of 
the ferromagnetic structure in each layer, 
while it does not contribute to interlayer singlet pairing. 
Therefore, the magnetic transition occurs at a higher temperature 
than the superconducting transition temperature 
unless the on-site $U$ is negligibly small.

Interlayer pairing has been studied 
by many authors~\cite{Kle91}. 
In this paper, 
we have examined the magnetic mechanism of the pairing interactions 
for interlayer singlet pairing, 
when the electrons are on the magnetic layers. 
However, irrespective of the pairing mechanism, 
interlayer pairing of the present type seems to be the most favorable, 
apart from equal spin pairing, 
when the present type of antiferromagnetic long-range order coexists. 
Other pairing states, such as intralayer singlet pairing, 
are strongly suppressed by the splitting of the Fermi-surfaces 
of the electrons with up and down spins 
due to the antiferromagnetic moment. 
Even in the two-fluid model, the exchange field must be 
induced on the electrons responsible to superconductivity.

In conclusion, in antiferromagnets with 
the magnetic order of the wave vector $\vQ = (0,0,\pi/c)$, 
magnetic interactions may induce the superconductivity 
of interlayer spin-singlet pairing, 
the order parameter of which has a horizontal line node. 
It was found that superconductivity and an antiferromagnetic 
long-range order with large localized magnetic moments $m$ 
can coexist, 
and that $\Tc$ is not influenced by the magnitude of $m$. 
It was also found that $\TAF > \Tc$ in most cases, 
unless $\TAF = 0$. 
The present model may describe an essential aspect 
of antiferromagnetic heavy fermion superconductors, 
such as ${\rm UPd_2Al_3}$ and ${\rm CePt_3Si}$. 
The orders of the magnitude of $\TAF$ and $\Tc$ and 
their ratio $\TAF/\Tc \approx 3 \sim 7$ can be reproduced 
by assuming moderate parameter values. 
The resultant order parameter is consistent with 
the observations mentioned above~\cite{Tou95,Ama92,Hes97,Wat04,Fey94,Iza05}.

The author wishes to thank Y.~Matsuda 
for giving their preprint on ${\rm UPd_2Al_3}$ 
and for useful discussions. 
This work was partially supported by 
the Ministry of Education, Science, Sports and Culture of Japan, 
Grant-in-Aid for Scientific Research (C), No.16540320, 2004.




\begin{thebibliography}{99}
\bibitem{NoteMiy01}
  The exchange interactions $J(\vq)$ considered in ref.~\cite{Miy01} 
  are those between the itinerant electrons and the localized spins 
  as $\JK$ in ref.~\cite{Shi94}, 
  while our ${\hat J}(\vq)$ in the present paper are those 
  between the itinerant electrons on adjacent layers.
\end{thebibliography}
\end{document}